\documentclass[prl,onecolumn,showpacs,preprintnumbers,amsmath,amssymb,floatfix]{revtex4-1}
\usepackage{graphicx}
\usepackage{dcolumn}
\usepackage{bm}
\usepackage[usenames]{color}

\usepackage[all]{xy}
\usepackage{mathtools}
\usepackage{amsfonts}
\usepackage{amssymb}
\usepackage{amsmath}
\usepackage{natbib}
\usepackage{longtable}
\def \bea{\begin{eqnarray}}
\def \eea{\end{eqnarray}}
\def \be{\begin{equation}}
\def \ee{\end{equation}}

\def\({\left(} \def\){\right)}

\DeclareMathOperator{\sech}{sech}
\begin{document}

\title{The Geometrical Structure of Bifurcations During Spatial Decision-Making}

\author{Dan Gorbonos$^{1}$,  Nir S. Gov$^{4}$ and Iain D. Couzin$^{1,2,3}$}

\affiliation{$^1$Department of Collective Behaviour, Max Planck Institute of Animal Behavior, 78464 Konstanz, Germany}
\affiliation{$^2$Centre for the Advanced Study of Collective Behaviour, University of Konstanz, 78464 Konstanz, Germany}
\affiliation{$^3$Department of Biology, University of Konstanz, 78464 Konstanz, Germany}
\affiliation{$^{4}$Department of Chemical and Biological Physics, Weizmann Institute of Science, Israel}

\begin{abstract}
Animals must constantly make decisions on the move, such as when choosing among multiple options, or ``targets'', in space. Recent evidence suggests that this results from a recursive feedback between the (vectorial) neural representation of the targets and the resulting motion defined by this consensus, which then changes the egocentric neural representation of the the options, and so on. Here we employ a simple model of this process to both explore how its dynamics account for the experimentally-observed abruptly-branching trajectories exhibited by animals during spatial decision-making, and to provide new insights into spatiotemporal computation. Essential neural dynamics, notably local excitation and long-range inhibition, are captured in our model via spin-system dynamics, with groups of Ising-spins representing neural ``activity bumps'' corresponding to target directions (as in a neural ring-attractor network, for example). Analysis, employing a novel ``mean-field trajectory'' approach, reveals the nature of the spontaneous symmetry breaking—bifurcations in the model that result in literal bifurcations in trajectory space---and how it results in new geometric principles for spatiotemporal decision-making. We find that all bifurcation points, beyond the very first, fall on a small number of ``bifurcation curves''. It is the spatial organization of these curves that is shown to be key to determining the shape of the trajectories, such as self-similar or space filling, exhibited during decision-making, irrespective of the trajectory’s starting point. Furthermore, we find that a non-Euclidean (neural) representation of space (effectively an elliptic geometry) considerably reduces the number of bifurcation points in many geometrical configurations (including from an infinite number to only three), preventing endless indecision and promoting effective spatial decision-making. This suggests that a non-Euclidean neural representation of space may be expected to have evolved across species in order to facilitate spatial decision-making.

\end{abstract}

\maketitle
\section{Introduction}

Selecting among spatially-discrete options (targets), such as food sources, shelters, or mates, is a ubiquitous challenge for animals. Despite this, very little research has been undertaken with respect to the mechanistic basis of spatial decision-making. For example, only recently have researchers explicitly considered the trajectories exhibited by animals when making such decisions~\cite{previous}. These experiments with both invertebrates (fruit flies and locusts) and a vertebrate (zebrafish) provided evidence that the brain repeatedly breaks multi-choice decisions into a series of binary decisions in space-time. This is reflected in abrupt changes in the trajectory of animals as they approach targets. We demonstrated, in a simple spin-based model of this cognitive process, that the bifurcations within the brain—that correspond to literal bifurcations in the trajectories (when summed over many repeated decisions)---is consistent with a recursive feedback between the (vectorial) neural representation of the different options and the animal’s movement. Spatial decision-making therefore appears to be an ‘embodied’ process that depends on the recurrent interplay between the time-varying egocentric neural representation of the options as animals move through space and the neural consensus dynamics that establish which direction to select at each moment in time. At such bifurcation points the spin model predicts that the brain spontaneously becomes extremely sensitive to very small differences between the options~\cite{previous}, which is a highly valuable property for decision-making.

Consistent with this model, the brains of a wide range of species have been shown to represent egocentric spatial relationships via explicit vectorial representation~\cite{Kim849,Seelig2015,Taube436,finkelstein2015three,Sarel176,hoydal2019object}. For example, neural ring-attractor models were motivated by such neuronal representations of the instantaneous heading direction of animals in the horizontal plane regardless their location and ongoing behavior~\cite{zhang,Touretzky}. When animals are exposed to a prominent target, an activity bump that corresponds to this target appears on a specified sector of the ring~\cite{Kim849,Seelig2015,Taube436,finkelstein2015three,Sarel176,hoydal2019object}, and this results with the animal turning towards it. After exposure to a new attractive landmark, the choice to turn to it is expressed as a shift in the activity bump towards the new target. The exact nature of the angular shift of the activity bump was found to be dependent on the relative angle of the new landmark from the older one~\cite{Kim849}. The interaction between the neurons within this network have been shown to exhibit local excitation (positive feedback among neural ensembles that represent directions with a small relative angle) and long range, or global, inhibition~\cite{zhang,Touretzky}(negative feedback between neural ensembles that represent directions with a large relative angle). This transition between types of feedback is responsible for the abrupt transition between two types of behavior - a movement in a compromise between the targets and a decision phase where the organism makes its way towards one of the targets. This behavior is captured in our spin-model - at a critical angle there exists a transition between a positive interaction between the spins (``ferromagnetic'') and a negative one (``antiferromagnetic''). By analogy, the spins in the model describe neuronal groups that either exhibit a relatively high, or low, firing rate, respectively. It was shown in~\cite{previous} that the number of spins that are active in a specific direction can be formally mapped to the firing rate of the corresponding neuronal cluster in the ring attractor class of models. 

The bifurcated trajectories observed for the fly, locust and fish~\cite{oscar2023simple}, when moving towards two or three targets, indicated that the spin-spin interactions in the model have an angular dependence that deviates from a simple vectorial dot product \cite{PhysRevE.97.032304}.
We therefore considered in~\cite{previous} a distortion of the angle in the
decision making process that represents neuronal encoding of space in a manner which could be non-Euclidean. This is supported by evidence for such distortion in the neuronal interactions~\cite{MULLER20051129,Sompolinsky}. 
The non-Euclidean distortion means that the egocentric representation of the targets (i.e., the geometry as represented by the neuronal activity) results in a ``perceived angle'' for frontal targets (with respect to the animal's heading direction) that is larger than the actual egocentric angle between them. Put another way, the brain ``expands'' the representation of space ahead of the animal. It is mediated by the transition of excitation to inhibition, which is a classic motif in the brain in general and in the ring attractor class of models in particular.


Despite the good agreement between the model and the experiments, many open questions remain. We introduced in~\cite{previous} a new tool to study the mean field solution which we called ``MF trajectory'', where we solved at each point in space the direction (or directions) of motion assuming that the system reached thermal equilibrium. The solution at each point gave us a vector (or vectors) that pointed towards the next point. This way we got a solution which is effectively at thermal equilibrium at each point in space. Using the MF trajectory we discovered the emergence of trajectories with an infinite series of bifurcations (for 3 targets), but it was not clear what determines this phenomena. In addition we found that the non-Euclidean distortion was roughly the same across species. These results naturally bring up the following questions: What (if anything) is special about this regime of distortion?
What controls the complexity, overall spatial organization and structure of the trajectories and their bifurcations?
Why do we find in the simulations (and consistent with experimental data) always two outgoing branches from each bifurcation?

These and other questions motivated us to study in greater detail the origin of the different classes of trajectories predicted by the model and the bifurcations along them. The analysis presented here exposes novel geometrical principles that relate the mutual arrangements of the bifurcations to the arrangement of the targets.

It is also interesting from a basic statistical physics point of view, as a new non-equilibrium system. Note that this is a complex physics problem, where the spin state defines the direction of motion, while the motion along the trajectory constantly changes the relative angles between the targets, as perceived by the moving animal, thereby changing the spin interactions (the Hamiltonian) along the trajectory. The present study of this problem also makes an advance in our understanding of this novel non-equilibrium statistical physics model, whereby the order parameter (“magnetization”) of the spins translates into direction of motion through space. Therefore it expands our understanding of a novel class of active complex systems, where self-propelled particles undergo internal ``decisions'' regarding their directions of motion.
This model also represents a new class of physical models by coupling spin-systems to movement dynamics, thus further combining 'classical' statistical physics with active matter research, thereby broadening the range of non-equilibrium physics.





\section{The spin model}

Here we generalize the spin model that was introduced in~\cite{PhysRevE.97.032304} to $k$ targets (see also in the SI of~\cite{previous}). The degrees of freedom are modeled as $N$ spins that are divided into $k$ equal subgroups. The direction of a target which is associated with a subgroup is denoted by $\hat{p}_i$ ($i=1,..,k$). Each individual spin in a subgroup has two states; ``on'', representing the neuronal firing ($\sigma_i=1$) or ``off'' ($\sigma_i=0$). The system can be described by the following Hamiltonian:
\be \label{Hamiltonian}
H=-\frac{k\,\bar{v}^2}{N}\sum_{i\neq j}\hat{p}_i\cdot\hat{p}_j\sigma_i\sigma_j,
\ee
where $\bar{v}$ is a constant whose dimension is velocity. The connectivity between the spins is all-to-all, for simplicity, which makes the mean-field (MF) solutions exact in the large $N$ limit.

The instantaneous velocity of the organism (or the group ~\cite{PhysRevE.97.032304}) is given by the sum of all the ``on'' spins in the direction of the respective targets 
\be
\vec{V}=\bar{v}\sum_{i=1}^{k}n_{i}\hat{p}_i, \label{velocity}
\ee
where $n_{i}=\frac{N_{i}}{N}$, $N_{i}$ is the number of individuals that exert a force in the group i (towards the target in the direction $\hat{p}_i$). When the dot product between two spins is positive ($\hat{p}_i\cdot\hat{p}_j>0$) the interaction is ferromagnetic (as between the spins in the same subgroup) and when it is negative, the interaction is antiferromagnetic. Below, we modify this dot product, motivated by recent models of neuronal perception~\cite{previous}.

The spin flip rates are constructed from the Hamiltonian (Eq.~(\ref{Hamiltonian})), in terms of Glauber dynamics~\cite{glauber1963time}:
\be \label{rates}
r^{(i)}_{1\rightarrow 0}=\frac{1}{1+\exp\(\frac{2\,k\,\bar{v}\,\vec{V}\cdot\hat{p}_i}{T}\)} 
 \quad\quad
r^{(i)}_{0\rightarrow 1}=\frac{1}{1+\exp\(-\frac{2\,k\,\bar{v}\,\vec{V}\cdot\hat{p}_i}{T}\)},
\ee
where $r^{(i)}_{1\rightarrow 0}$ is the rate in which a spin in group $i$ is switched off and
$r^{(i)}_{0\rightarrow 1}$ is the rate in which the spin is switched on. The temperature in our model describes the noise that drives random spin-flipping dynamics ~\cite{PhysRevE.97.032304}. Within the context of neuronal dynamics, the temperature $T$ represents the stochastic noise of neuronal firing. The equations of motion for the number of active spins in each subgroup (master equation), in the limit of $N\gg1$, are:
\be \label{MasterOld}
\frac{d\,n_{i}}{dt}=\frac{\frac{1}{k}-n_i}{1+\exp\(-\frac{2\,k\,\bar{v}\,\vec{V}\cdot\hat{p}_i}{T}\)}-\frac{n_i}{1+\exp\(\frac{2\,k\,\bar{v}\,\vec{V}\cdot\hat{p}_i}{T}\)}.
\ee

It is convenient to rearrange Eq.~(\ref{MasterOld}) to get
\be \label{Master}
\frac{d\,n_{i}}{dt}=\frac{1}{k\(1+\exp\(-\frac{2\,k\,\bar{v}\,\vec{V}\cdot\hat{p}_i}{T}\)\)}-n_i.
\ee
The steady-state solution of Eq.~(\ref{Master}) can be written as the solution of the following system of algebraic equations:
\be \label{sol}
n_i=\frac{1}{k\(1+\exp\(-\frac{2\,k\,\bar{v}\,\vec{V}\cdot\hat{p}_i}{T}\)\)} \quad\quad\quad i=1,...,k.
\ee
The system of MF equations include the $k+2$ equations (\ref{velocity}) and (\ref{sol}), whose solutions are the steady-state velocity and the fraction of active spins in each group: ($\vec{V}_{ss},n_{i,ss}$) ($i=1,..,k$). We will refer to this system as the MF steady-state (MFSS) solutions. 

\section{Mean Field trajectories and bifurcation points}

Here we introduce ``MF trajectories'' as a new tool to study the dynamics, while in ~\cite{PhysRevE.97.032304,previous} the trajectories were found by simulations. The nature of these trajectories, and the types of bifurcations along them, are explored below for systems with two or three targets.

\subsection{Two targets}
\begin{figure}[htb]
\centering
\includegraphics[width=0.5\linewidth]{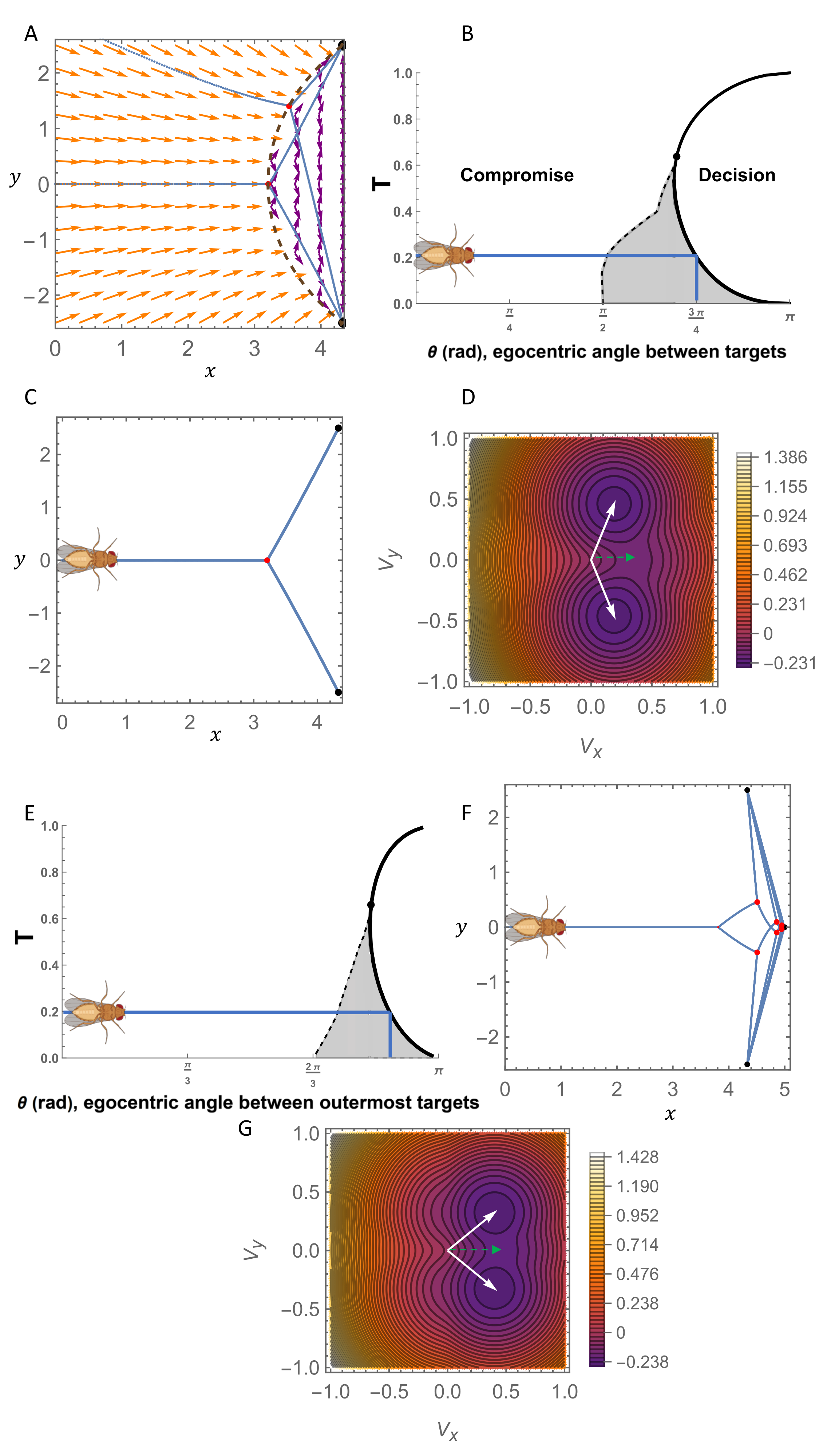}
\caption{\label{PhaseDiagram}The phase transition for two and three targets. 
(A) Vector fields that represent the stable solutions for the velocity at each point for two targets. The curve of first bifurcation is given in dashed brown, where the velocity in the original direction of motion becomes unstable. It separates between the regime where compromise stable solutions exist (whose vector fields are given in orange), and the regime of decision (stable solutions in purple arrows). Examples of two possible trajectories are given in blue.  (B) A phase diagram describing the phase transition exhibited in a bifurcation in space while moving from compromise to decision between two targets. The shaded area (also in E) represents the region in parameter space where both the compromise and the decision solutions exist. The dashed black line is the coexistence curve (binodal) where the additional phase (the decision) appears as an additional stable solution. The solid black line is the spinodal curve where the compromise solution becomes unstable and the system has to move into a stable decision phase (towards one of the two targets). The black dot is the tricritical point that separates the second order phase transition (above) from the first order (below). The critical angle at $T=0.2$ (between the two targets) corresponds to the MF trajectory in C. (C) The trajectory in space with the bifurcation point that corresponds to a phase transition on the spinodal curve at $T=0.2$ as shown in the phase diagram in B. The bifurcation point is marked by a red dot. The targets are at $(4.33,2.5)$ and $(4.33,-2.5)$ (as in~\cite{previous}) (D) The effective free energy landscape at the bifurcation point of C as a function of the velocity $\vec{V}=(V_x,V_y)$. At the phase transition point the velocity in the original direction becomes an unstable saddle point (green arrow) and the two new stable solutions (white arrows) correspond the two new minima (spontaneous symmetry breaking). 
(E) A phase diagram describing the first bifurcation in the case of three targets. The angle is measured between the external targets (the left and the right targets).  The critical angle at $T=0.2$ corresponds to the mean field trajectory in F. (F) The spatial trajectory in the MF approximation that corresponds to a phase transition on the spinodal curve at $T=0.2$ as shown in the phase diagram in E. The targets (black dots) are at $(4.33,2.5)$, $(4.33,-2.5)$ and $(5,0)$ (as in~\cite{previous}). The bifurcation points are marked by red dots. The emerging pattern is explained later in Fig. ~\ref{BifurCurves}. Only the first 4 bifurcation points are shown in every path (``bifurcation depth'' of 4) (G)  The effective free energy landscape at the first bifurcation point in F as a function of the velocity $\vec{V}=(V_x,V_y)$. (similar to D).  }
\end{figure}

At each point in space we can calculate the stable steady-state solutions ($\vec{V},n_i$), which give the MF direction vectors that are possible at that point. This defines a global flow field (Fig.~\ref{PhaseDiagram}A) along which the animal moves, in the limit of slow speed, such that the spins have time at each position to equilibrate to the MFSS solutions, and the trajectory follows the flow field defined by these stable solutions. For targets that are at infinity, the angles between them are constant, and the Hamiltonian (Eq.~(\ref{Hamiltonian})) is time-independent. However, for targets at finite distances, the relative angles change along the trajectory. Furthermore, the animal can move from a region with a single solution to a region where there are several stable solutions, and there is a spontaneous symmetry breaking. The location along the trajectory where such a transition occurs is defined as a bifurcation point, as the trajectory splits into several possible trajectories that emerge from that point.


We can define a bifurcation point where the MFSS solution along which the animal was moving, becomes unstable, and write the criterion for instability of a general MFSS solution. The dynamical equation for the velocity can be obtained using equations ($\ref{velocity}$) and ($\ref{Master}$):
\be \label{DynVel}
\frac{d\vec{V}}{dt}=\bar{v}\sum^{k}_{i=1}\frac{d n_i}{dt}\hat{p}_i=\sum^{k}_{i=1}\frac{\bar{v}}{k\(1+\exp\(-\frac{2\,k\,\bar{v}\,\vec{V}\cdot\hat{p}_i}{T}\)\)}\hat{p}_i-\vec{V}
\ee

Let $\vec{V}_{ss}$ be a solution of the model equations and consider a small perturbation to the velocity in the perpendicular direction to the velocity $\vec{V}$:
\be
\vec{V}=\vec{V}_{ss}+\epsilon\hat{n}
\ee
where $\hat{n}$  is the normal to $\vec{V}$. Substituting into Eq.~(\ref{DynVel}), expanding to first order in $\epsilon$ and taking the normal component we get the following equation for the perturbation $\epsilon$
\be
\frac{d\epsilon}{dt}=-A\epsilon+\mathcal{O}\(\epsilon^2\)
\ee
where
\be \label{Adefined}
A\equiv 1-\frac{\bar{v}^2}{2\,T}\sum^{k}_{i=1}\sech^2\(\frac{k\,\bar{v}\,\vec{V}\cdot \hat{p}_i}{T}\)(\hat{n}\cdot \hat{p}_i)^2.
\ee
Then the solution $\vec{V}$ is stable if $A>0$ and unstable if $A<0$. Therefore the bifurcation occurs where $A=0$.

This definition means that we choose to construct the MF trajectories such that the bifurcation occurs at the spinodal curve, according to the MF phase diagram, as shown for two targets in Fig.~\ref{PhaseDiagram}B. The transition could be chosen to occur at any point in the hysteresis region of the phase diagram (grey region), between the binodal (where two new stable solutions first appear) and spinodal (where the original ``compromise'' direction becomes unstable) lines. We show in the appendix that at the point where a MF solution becomes unstable, there are at least two alternative stable solutions. Then we repeat the same procedure along the new stable solutions and this way form a directed graph whose nodes are the bifurcation points. 

The MF trajectory approaching symmetrically two targets is given in Fig.~$\ref{PhaseDiagram}$B for $T=0.2$ (the geometrical arrangement of the targets as in the experimental setup, see Fig.~1 in ~\cite{previous}). The critical bifurcation angle between the targets, where the bifurcation point occurs, is the one that corresponds to the intersection of the spinodal curve and the horizontal line $T=0.2$ in the phase diagram (Fig.~\ref{PhaseDiagram}B). At higher temperatures, above the tri-critical point where the binodal and spinodal curves coincide, the transition is second order and there is no hysteresis region. 


Another way to examine the bifurcation points is using the effective free energy (whose derivation appears in the appendix, and in~\cite{PhysRevE.97.032304}), which is given by 
\be
F(\vec{V},T)=k\,\vec{V}^2-\frac{T}{k}\sum_{i=1}^{k}\ln\(1+\exp\(\frac{2\,k\,\bar{v}}{T}\,\vec{V}\cdot\widehat{p}_i\)\). \label{free}
\ee

The effective free energy at a specific point in space is a function of the velocity and the angles towards the targets, and it shows the energetic cost of different spin configurations that correspond to different velocities (the ``free energy landscape''). The stable MFSS solutions correspond to minima of this free energy. In Fig.~$\ref{PhaseDiagram}$D we see the free energy function at the bifurcation point of Fig.~$\ref{PhaseDiagram}$C, 
where the original direction of movement becomes a saddle point, and the two new minima correspond to the two new stable solutions and directions of motion.

In Fig.~\ref{PhaseDiagram}A we plot the flow field, of the MFSS solutions. The bifurcation points for two different trajectories are denoted, lying along a curve that connects the two targets. We find this curve by requiring that the compromise between the two targets $n_1=n_2\equiv n$ is a stable solution (Eqs.~($\ref{velocity}$,$\ref{sol}$)), while requiring that $A=0$ at the bifurcation points (Eq.~($\ref{Adefined}$)). We then obtain the following two equations:
\bea \label{compromise2}
\frac{2\,T}{\bar{v}^2}&=&\sech^2\(\frac{2\,\bar{v}\,V_{p}}{T}\)(\sin^2\theta_1+\sin^2\theta_2)\\
n&=&\frac{1}{2\(1+\exp\(-\frac{4\,\bar{v}\,V_{p}}{T}\)\)}
\eea
where 
\be
V_{p}=\bar{v}\,n\(1+\cos(\theta_1+\theta_2)\)
\ee
and we denote by $\theta_{1,2}$ the angles to the respective targets as measured relative to the $x$ axis. We have three variables ($\theta_1,\theta_2,n$) for two equations and therefore a one parameter family of solutions which defines the bifurcation curve that connects the two targets (the dashed brown line in Fig.~\ref{PhaseDiagram}A).
Using the effective free energy we can also prove that the description of a bifurcation point as a point of instability of one solution is equivalent to the existence of at least two stable solutions at this point. The proof appears in the appendix.

\subsection{Three targets}

In the case of three targets there is more than one bifurcation point along the trajectory. Let's consider a symmetric starting point, as shown in Fig.~\ref{PhaseDiagram}F. The first bifurcation is between a compromise of all the three targets (for $\theta<\theta_c$) and a compromise of only two targets (while the remaining target is suppressed). The phase diagram for this first bifurcation point, along a symmetric trajectory, is given in Fig.~$\ref{PhaseDiagram}$E, with the corresponding MF trajectory in Fig.~$\ref{PhaseDiagram}$F. The effective energy landscape for this first bifurcation is shown in Fig.~\ref{PhaseDiagram}G, and has the same binary structure as for the bifurcation of two targets (Fig.~\ref{PhaseDiagram}D).

At the second bifurcation we find that the trajectories become much more complex (Fig.~\ref{PhaseDiagram}F), with one solution pointing towards an edge target, while the second solution is a compromise of the remaining two targets. Subsequently there are more bifurcations of the same type, alternating between the outermost targets and compromise solutions (between the left and central targets, or between the right and central targets). This pattern introduces an infinite number of bifurcations that become infinitely dense towards the central target. In the appendix (Fig.~\ref{Convergence}) we show numerically that this sequence converges into an infinite geometric series towards the central target, where the pattern becomes self-similar. For this particular arrangement of targets, the MF trajectories are found to reach the central target after an infinite series of bifurcations. The presence of noise during the motion of a real animal, or during dynamic simulations, enables trajectories to reach the central target ~\cite{previous}.

\begin{figure}[htb]
\centering
\includegraphics[width=0.83\linewidth]{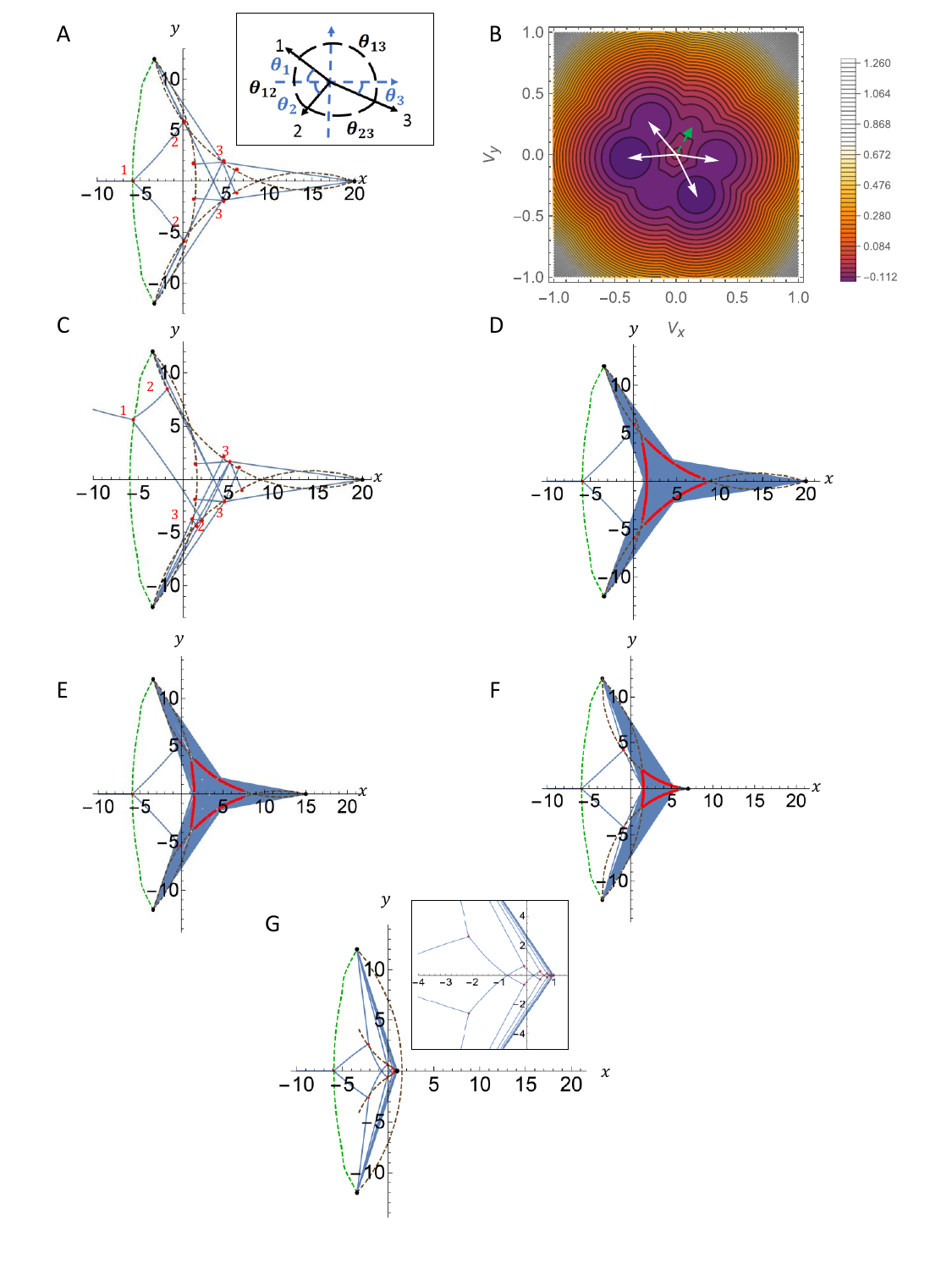}
\caption{\label{BifurCurves}
Bifurcation Curves. 
(A) A trajectory that starts from $(-15,0)$ towards three targets at $(-3.4,12)$, $(-3.4,-12)$ and $(20,0)$ (shown only starting from $x=-10$). The depth is 4 bifurcation points. Bifurcation points are in red and bifurcation curves are the dashed brown curves. A curve of first bifurcation is in dashed green. The first three bifurcation points are numbered by their order. The inset shows the definition of the angles that are used in the parametrization of the bifurcation curves for three targets that correspond to compromise of two of the three targets (the dashed brown curves). (B) The energy diagram at the third bifurcation point of the diagram in A. The original direction becomes unstable (dashed green) and there are four minima that correspond to the four possible directions (solid white). (C) The same like A (four first bifurcation points) but with asymmetric initial conditions (starting from $(-15,8)$). The trajectory still converges towards the same bifurcation curves. (D)-(G) 
The formation of a limiting self-similar structure as a result of the absence of one of the bifurcation curves at $T=0.2$. The bifurcation points are in red and the bifurcation curves are in dashed brown. The depth is 12 bifurcation points.  The starting point of all the trajectories is at (-10,0). Two of the targets are at (-3.4,12) and (-3.4,-12). The only difference between the four plots is in the position of the central target.
(D)  The central target is at (20,0).  Note how the bifurcation points are dense (in a segment of the bifurcation
curve).
(E)  The central target is at (15,0)
(F) The central target is at (7,0)
(G) The central target is at (1,0). Here the target is before the bifurcation curve that connects the left and right targets, and thus a self-similar pattern emerges (in the limit). The inset shows a magnification of the pattern.}
\end{figure}

The number and positions of the bifurcation points strongly depends on the geometrical arrangement of the targets. In Fig.~\ref{BifurCurves}A-D we demonstrate this by calculating the trajectories for a geometry where the central target is shifted to be further behind the two edge targets, compared to the arrangement of Fig.~\ref{PhaseDiagram}F. In this configuration, the trajectories are not self-similar, and there are trajectories that reach the central target, as shown in Fig.~\ref{BifurCurves}B,C by following the trajectories up to the fourth bifurcations. In the case of three targets, the curve on which the first bifurcation points lie does not correspond to a compromise between targets, and unlike the case of two targets we do not have an analytic condition for this curve. It is found numerically, denoted by the green dashed line in Fig.~\ref{BifurCurves} for the trajectories that start to the left of the edge targets.

The second bifurcation along the trajectories is also binary, leading to one of the edge targets or to a compromise of the remaining targets. However, the third bifurcation is sometimes split into more than two outgoing trajectories (Fig.~\ref{BifurCurves}B,C). In general, the trajectory arriving at the third bifurcation was a compromise between two targets. This compromise becomes unstable at the ``bifurcation curves'' (BC) denoted by the dashed brown lines in Fig.~\ref{BifurCurves}. Therefore, from each bifurcation point on these BC, there can be up to four outgoing trajectories (Figs.~\ref{BifurCurves}B,C): two ``decision'' trajectories towards the two targets that were in compromise when arriving at the BC, and two ``compromise'' solutions that involve the third target.
Please note, however, as will be shown using dynamic simulations in the following section, despite a larger possible number of possible outgoing trajectories in these cases, we find that in the presence of stochasticity only two actually occur at each bifurcation point.



The BC, and their organization with respect to the targets, are key to determining the shape of the trajectories, as all the bifurcation points of second and higher order reside on them. Let us start by specifying the analytic conditions that these BC obey: We require an instability $A=0$ (Eq.~(\ref{Adefined})), in which there is a compromise of two group of spins (in the case of three targets, in analogy to Eq.~(\ref{compromise2}) for two targets). By requiring a compromise of two of the three targets, we obtain three pairs that give us the three BC in Fig~\ref{BifurCurves}B,C. Let us consider for example the trajectories where we have compromise between two targets ($n_1=n_2$). It's convenient to parameterize the points according to three angles that are given in the inset of Fig.~\ref{BifurCurves}A. For a given point $\theta_1,\theta_2,\theta_3$ are angles to the corresponding target directions $1,2,3$ and in addition we define the following angles: 
\bea
\theta_{12}&=&\theta_1+\theta_2, \nonumber \\
\theta_{23}&=&\pi-\theta_2-\theta_3, \nonumber\\
\theta_{13}&=&\pi-\theta_1+\theta_3,
\eea
that we substitute into the velocity components, that are given for the case of three targets in Eqs.~(\ref{vp1},\ref{vp2},\ref{vp3}), to give us 
\bea 
V_{p1}&=&\bar{v}
\(n_1+n_2\cos(\theta_1+\theta_2)-n_3\cos(\theta_1-\theta_3)\),\\
V_{p2}&=&\bar{v}\(n_1\cos(\theta_1+\theta_2)+n_2-n_3\cos(\theta_2+\theta_3)\), \\
V_{p3}&=&\bar{v}\(n_3-n_1\cos(\theta_1-\theta_3)-n_2\cos(\theta_2+\theta_3)\).
\eea

Substituting into the steady state equations~\ref{ni3} and~\ref{stability}, where we take $A=0$ and compromise between two targets at the bifurcation points, we get the following system of five equations:
\bea
n_1&=&n_2,\\
\frac{2\,T}{\bar{v}^2}&=&\sech^2\(\frac{3\,\bar{v}\,V_{p_1}}{T}\)\sin^2\theta_1+\sech^2\(\frac{3\,\bar{v}\,V_{p_2}}{T}\)\sin^2\theta_2+\sech^2\(\frac{3\,\bar{v}\,V_{p_3}}{T}\)\sin^2\theta_3,\label{StabThre}\\
n_1&=&\frac{1}{3\(1+\exp\(-\frac{6\,\bar{v}\,V_{p_1}}{T}\)\)},\\
n_2&=&\frac{1}{3\(1+\exp\(-\frac{6\,\bar{v}\,V_{p_2}}{T}\)\)},\\
n_3&=&\frac{1}{3\(1+\exp\(-\frac{6\,\bar{v}\,V_{p_3}}{T}\)\)},
\eea
for six variables: $n_1,n_2,n_3,\theta_1,\theta_2,\theta_3$. Then the solution can be parameterized by one parameter that gives us a curve that connects pairs of targets (Fig.~\ref{BifurCurves}B,C dashed brown curves). In Fig.~\ref{BifurCurves}B,C we show the first four bifurcation points, which lie on the BC. In Fig~\ref{BifurCurves}D we follow the trajectory up to high order (12) of bifurcations, and we find that the trajectory has a space-filling property, with the bifurcation points accumulating along the central part of the BC.

Furthermore, the BC are determined by the relative positions of the targets and not by the starting point of the trajectory (Fig.~\ref{BifurCurves}B,C). In this sense the BC act as attractors of the MF trajectories. We demonstrate this by plotting the trajectories and the BC for configurations where the central target is shifted closer to the edge targets, as shown in Fig.~\ref{BifurCurves}E-G by following the trajectories up to 12th bifurcation order. Since the edge targets remain in the same positions as in Figs.~\ref{BifurCurves}B-D, the BC that connects these two targets remains the same. By shifting the central target towards the edge targets we reach a configuration (Fig.~\ref{BifurCurves}G) where this BC is on the other side of the central target, and thus cannot be reached by trajectories that bifurcate along the other two BC. As a result, we obtain only two reachable BC (dashed brown lines in Fig.~\ref{BifurCurves}G), and an infinite interplay between them which converges to a self similar pattern (Fig.\ref{PhaseDiagram}F, and see the appendix for more details). 

The dependence of the bifurcation structure on the temperature is given in appendix (Figs.~\ref{Tt0},~\ref{SecondBifurcationTgg0},~\ref{HighTexamples}). In particular, we find that the BC simplify significantly and become straight lines between the targets in the limit of $T\rightarrow 0$.

\section{Comparison between simulated and MF trajectories}

Since real animals and groups move at a finite speed, and are exposed to sources of noise (e.g., sensory noise), we next explore the relation between the deterministic MF trajectories and noisy simulations of the spatial decision-making process. In \cite{previous} we showed that these sources of noise can smear out the fractal-like trajectories predicted by the MF (Fig.~\ref{BifurCurves}G), in the numerical simulations. 

\begin{figure}
\centering
\includegraphics[width=0.95\linewidth]{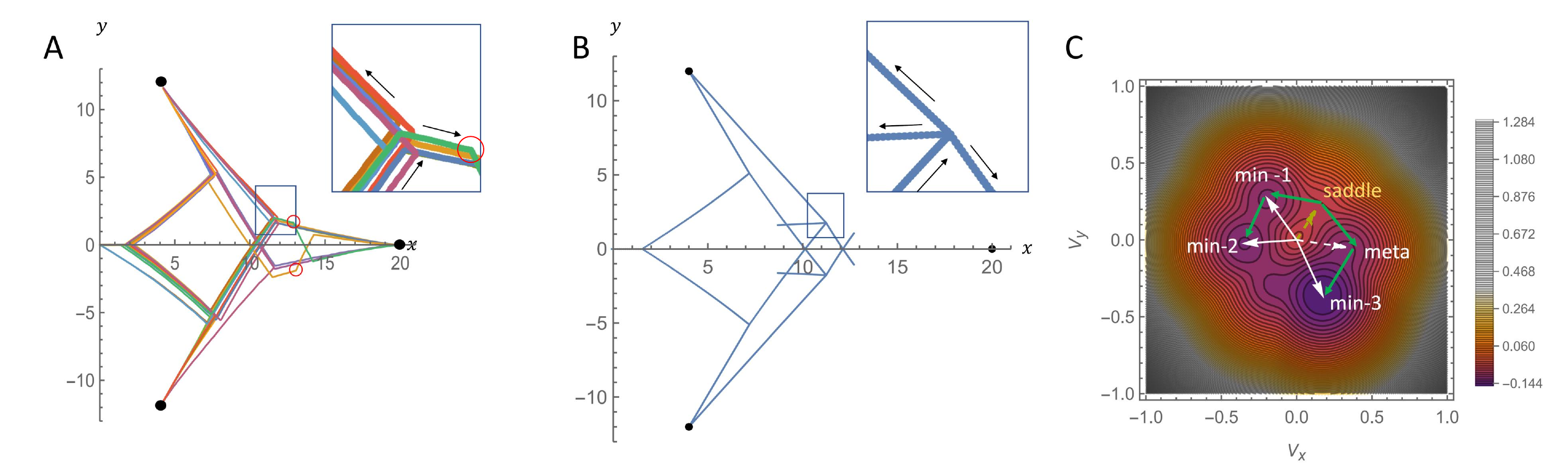}
\caption{\label{Simulation}
The simulation.
(A) Some trajectories from the simulation, each trajectory is in a different color.
Starting point is $(0,0)$. The targets are at $(4,12)$,$(4,-12)$ and $(20,0)$ (in black). We want here to examine the the trajectories at the third bifurcation point (marked by the blue square). The red circles denote sharp turning points of trajectories in which saddle points (of the effective free energy) evolve into a local minima.
(B) The MF trajectory until the third bifurcation point, given here for comparison. The three outgoing lines correspond to the three minima at this point.
(C) The energy diagram at the third bifurcation point. The original direction before the bifurcation (yellow arrow) becomes a saddle point, and thus unstable. The diffusion to the new possible solutions is denoted by the green arrows. There is one meta-stable solution and three minima that correspond to the three outgoing directions in B (the MF). Highest probability is to end up in one of the two solutions that are closest to the original unstable one. It explains why the bifurcations in the simulation are always binary without trajectories directed backwards.   
}
\end{figure}

In the numerical simulations (Fig.~\ref{Simulation}) each target is represented by a small number of spins (15 per target), and therefore has intrinsic finite-size noise. In addition, the position of the moving animal (or group) is updated with a time step that allows for about 10 spin updates, using Glauber dynamics (Eq.~(\ref{rates})). This procedure of finite time per movement step is different from the MF behavior, which is the limit of infinitely slow movements. Unlike the MF calculation where the spins are at their equilibrium at each point along the trajectory, the spins are not at equilibrium along the simulated trajectory. Thus, the simulated trajectories exhibit bifurcations that are not always exactly at the predicted locations in the MF analysis. However, the general structure of the trajectories and the outgoing directions from each bifurcation, are in good similarity with the MF result. 


Examples of simulation trajectories are given in Fig.~\ref{Simulation}A, which are compared to the corresponding MF trajectory (Fig.~\ref{Simulation}B). Let us focus on the third bifurcation, denoted by the blue frame in Fig.~\ref{Simulation}A,B, and magnified in the insets. In this simulated bifurcation there are only two outgoing trajectories, while in the MF trajectory we find three outgoing trajectories and in different directions. We can explain these differences by looking at the energy lanscape for this bifurcation, in Fig.~\ref{Simulation}C. The green arrow indicates the original unstable direction. The green arrows show the time evolution of the spins, as the system flows towards the newly available minima. The MF directions correspond to the three new minima at this bifurcation point, while the directions in which the system leaves the bifurcation in the simulations correspond to the minimum (``min-1'') and saddle point (``meta'') that are closest to the original direction. Due to the short time available for the spins to evolve at each point along the simulated trajectory, the system moves along the two directions where the spin evolution slows down, corresponding to the the minima or saddle point closest to the original direction before the bifurcation point. Trajectories pointing towards the compromise of the edge targets (``min-2'' in Fig.~\ref{Simulation}C) are therefore not observed in the simulations. When the system continues along the saddle-point direction (``meta'' in Fig.~\ref{Simulation}C), it sometimes becomes a true minimum after a short time, allowing the system to continue towards the central target. This way, in this example, the system can reach the central target from the third bifurcation point, even though we do not see it in the MF trajectory. In other cases, we see that the spins evolve away from the saddle to the local minimum (``min-3'' in Fig.~\ref{Simulation}C), with a corresponding sharp turn in the trajectories (denoted by red circles in Fig.~\ref{Simulation}A).   

This example explains the principle as to why we never see bifurcations that are non-binary in the simulations and in the experimental data \cite{previous}, despite the existence of such complex bifurcations in the MF trajectories.

\begin{figure}
\centering
\includegraphics[width=0.75\linewidth]{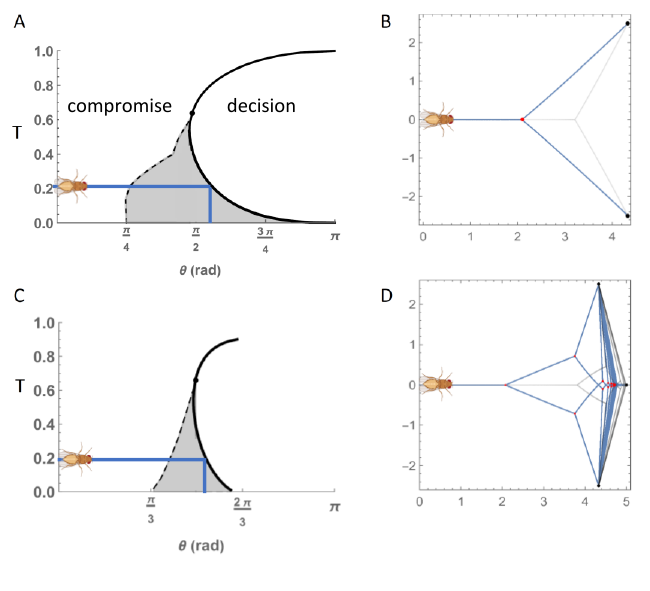}
\caption{\label{PhaseDiagramnu}
The phase transition at $\nu=0.5$, which is the value of the angular distortion according to the experimental results in~\cite{previous}. (A) A phase diagram describing the phase transition exhibited in a bifurcation in space while moving from compromise to decision between two targets at $\nu=0.5$. (B) The MF trajectory (in blue) for two targets with the bifurcation point that corresponds to a phase transition on the spinodal curve at $T=0.2$ as shown in the phase diagram in A. The bifurcation point is marked by a red dot. The targets are at $(4.33,2.5)$ and $(4.33,-2.5)$ (as in~\cite{previous}). Note that the bifurcation here happens at a smaller critical angle compared to $\nu=1$ (Given here in light gray for comparison, see Fig. \ref{PhaseDiagram}) (C) A phase diagram describing the first bifurcation in the case of three targets at $\nu=0.5$. (D) The MF trajectory (in blue) for three targets with the bifurcation point that corresponds to a phase transition on the spinodal curve at $T=0.2$ as shown in the phase diagram in C. The targets (black dots) are at $(4.33,2.5)$, $(4.33,-2.5)$ and $(5,0)$ like in Fig.~\ref{PhaseDiagram}F. Since the bifurcations (the red dots) appear in smaller critical angles the self similar pattern, which is explained in Fig.~\ref{BifurCurves}), is shown more explicitly compared to $\nu=1$ in Fig.~\ref{PhaseDiagram}F (which is given here for comparison in light gray). Note that in principle it should continue until the central target at $(5,0)$ and only the first 12 bifurcations are shown here (the ``bifurcation depth'' is 12).
 }
\end{figure}

\begin{figure}[htb]
\centering
\includegraphics[width=0.75\linewidth]{
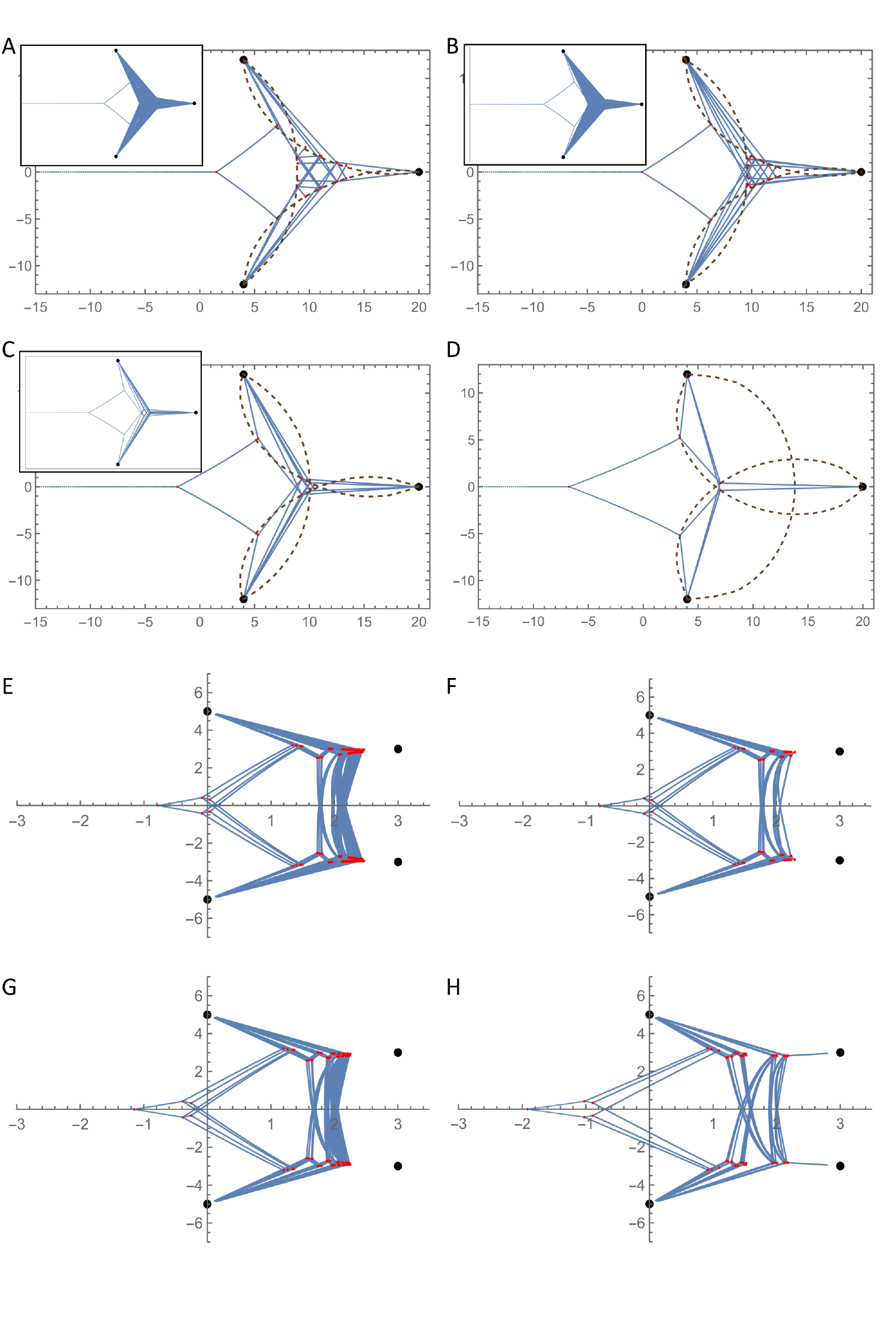}
\caption{\label{Reducing}
An example for the reduction of structure complexity of trajectories when the angular distortion is increased (smaller $\nu$). The targets are at $(4,12)$ - left,$(4,-12)$ - right and $(20,0)$ - central.  A depth of five bifurcation points is given here. $T=0.2$.
(A) $\nu=1$
(B) $\nu=0.85$
(C) $\nu=0.7$
(D) $\nu=0.5$.
Bifurcation curves are in dashed brown. In the insets we see the same trajectories with depth of 12 bifurcation points that show how the space filling area shrinks as $\nu$ is reduced. In the case of $\nu=0.5$, the total number of bifurcations is three (and therefore no inset).
In the case of four targets loops might be formed and smaller values of $\nu$ (stronger angular distortion) help to escape the loops. The targets (in black) are at $(0,5)$,$(3,3)$,$(3,-3)$ and $(0,-5)$. The temperature is always $T=0.2$, and the trajectory starts from $(-2,0)$. The bifurcation points are in red.
(E) $\bm{\nu=1}$ - no angular distortion. Depth of 12 bifurcation points.  The loops do not overlap each other but remain tight.
(F) $\bm{\nu=1}$ - no angular distortion. Depth of 9 bifurcation points.  Here only the first few loops are shown. Note how the loops are formed out of a series of bifurcations.
(G)$\bm{\nu=0.9}$. Depth of 9 bifurcation points.  Due to the angular distortion the bifurcations appear earlier along the trajectories.
(H) $\bm{\nu=0.75}$. Depth of 8 bifurcation points.  In addition to earlier bifurcations, the angular distortion is so strong that direct trajectories are formed towards the targets on the right - a way out of the loop.The aspect ratio in figures E-H is not equal to 1 in order to show the loop structure clearly. 
}
\end{figure}

\section{The non-Euclidean Encoding of Space}

Another aspect of the trajectories that we found when comparing our model to the experiments \cite{previous}, is that the interactions between the spins are modified from the simple cosine function of the relative angles (dot product in Eq.~(\ref{Hamiltonian})). We now explore how this modified interaction affects the MF trajectories, which can shed light on the features that this modification provides. 

The angle between targets at $\hat{p}_i$ and $\hat{p}_j$, $\theta_{ij}$, denote the angle between targets at $\hat{p}_i$ and $\hat{p}_j$, so that
\be \label{theta_def}
\cos\theta_{ij}=\widehat{p}_i\cdot\widehat{p}_j
\ee
The modified interaction between the spins that represent targets $i,j$ is simply the cosine of the distorted angle $\tilde{\theta}_{ij}$, parameterized by $\nu$
\be
\theta_{ij} \rightarrow \tilde{\theta}_{ij}=\pi\,\(\frac{\theta_{ij}}{\pi}\)^{\nu}. \label{dist}
\ee


For $\nu<1$ in Eq.~(\ref{dist}) we get a distortion of the Euclidean angle between the directions of the targets corresponding to the interaction becoming negative at a smaller relative angle, and as a result a non-Euclidean encoding of space (effectively an elliptic geometry). It was shown in~\cite{previous} that 
comparing to the experimental data this parameter appears to exhibit variability within species.
From comparison to a series of experiments in fruit flies, locusts and zebrafish a common value of $\nu\sim 0.5$ was obtained for the majority of the individual animals, while there are individuals 
that behave in a way consistent with having a smaller value of this parameter, and therefore their trajectories are more direct with respect to their selected target and tend not to exhibit bifurcations (this increases the speed of decision making, but at the cost of decreased discriminatory capability when targets differ in quality). 

Here we plot the phase diagrams for  $\nu\sim 0.5$ and a symmetric trajectory towards two and three targets (Fig.~\ref{PhaseDiagramnu}A,C), which shows that the transition lines are shifted to lower angles compared to Fig.~\ref{PhaseDiagram}B,E. The corresponding MF trajectories for the two and three target systems are shown in Fig.~\ref{PhaseDiagramnu}B,D, compared to the trajectories for simple cosine interactions ($\nu=1$, Fig.~\ref{PhaseDiagram}C,F).
The stability test in Eq.~(\ref{Adefined}), which we used before to identify the bifurcations along the MF trajectories, cannot be generalized to $\nu<1$ directly. For this purpose we had to rewrite it into an equivalent form which is given in the appendix (Eq.~(\ref{Stab_alt})). Using this equivalent form, we can also obtain equations for the BC, for the modified interactions for $\nu<1$,  by replacing Eq.~(\ref{StabThre}) with Eq.~(\ref{Stab_alt}). 

Next we want to examine how the trajectories change when the angular distortion is introduced, namely as $\nu$ is decreased below $1$.
For the case of three targets, we find that for decreasing $\nu$ there is a reduction in the complexity of the trajectories, as shown for example in Fig.~\ref{Reducing}A-D. Due to the reduction in $\nu$ the bifurcations happen earlier along the trajectory at a smaller relative angle (as in Fig.~\ref{PhaseDiagramnu}), and this way many complexities of the bifurcation pattern shrink and sometimes vanish. The number of bifurcations is reduced from infinity for $\nu=1$ up to three bifurcations for $\nu \lesssim 0.55$. 

In the case of four targets we find that the effects of the angular distortion can be even more dramatic. For the regular cosine ($\nu=1$) interactions, we find loops in the trajectories  (Fig.~\ref{Reducing}E), which form due to an endless series of bifurcations, and do not allow any trajectory to reach the middle targets (the two targets that are on the right side). We show in detail how loops are formed in this case in the appendix. Applying angular distortion (Fig.~\ref{Reducing}F-H) we find that the whole bifurcation tree becomes more spatially spread, and trajectories that lead to the middle targets appear. 
Since in~\cite{previous,oscar2023simple} we found evidence that animals seem to have $\nu\sim 0.5$, it may well be that such loops are very rare in nature. Our result suggests that animals evolved to avoid such detrimental behavior by having a narrow attention field (small $\nu$). In addition, sufficient noise will also tend to resolve such endless loops, thereby also limiting the possibility of observing this behavior.

Overall, the MF trajectories in Figs.~\ref{PhaseDiagramnu}-~\ref{Reducing} demonstrate that by employing interactions based on the distorted relative angle, with $\nu$ significantly lower than $1$, the animals (and groups) achieve trajectories that have bifurcations that are spread over a much larger spatial domain. This spreading of the bifurcation network simplifies the trajectories, and resolves pathological paths such as loops that never reach certain targets. This leads in reality to a more uniform probability of reaching the different targets, thereby making foraging more efficient and less prone to biases based on the geometry of the targets' relative locations.

\section{Discussion}

We presented here a detailed theoretical exploration of the mean field (MF) trajectories that arise from our spin-based decision-making model, which describes how single animals (and perhaps also animal groups~\cite{previous}) navigate towards an array of identical targets. This model was recently shown to predict tree-like trajectories with bifurcations, that were found in experiments and simulations of single animal and group movement \cite{previous}. Here we showed that the bifurcations along these trajectories lie along ``bifurcation curves'' (BC), and the organization of these BC determine the global nature of the trajectories. These BC are dependent on the spatial organization of the targets, and thus do not depend on specific initial conditions of the trajectory. The BC serve as attractors of the bifurcation points, such that any trajectory tends to converge to the same pattern of transitional movement between the BC. We show that the mutual arrangement of BC determine the qualitative nature of the trajectories. For example, for three targets, if all the three BC intersect each other we get a space filling trajectory, and if they do not intersect each other we get a self-similar trajectory.

The MF analysis of the trajectories also allows us to map the energy landscape in the spin/velocity space at the bifurcation points, thereby exposing when the outgoing paths will reach directly one of the targets, or continue along other compromise directions between targets. We can explain why we see in simulations only two branches (two outgoing trajectories) at every bifurcation point - at the bifurcation point the direction of motion shifts towards the new stable directions of motion, ending up at the first minima or saddle point (``exit point'') that are closest to the original direction of movement (as explained in Fig.~\ref{Simulation}).

In \cite{previous} it was shown that the sensitivity of the animal to external cues is maximal when it is moving through a bifurcation, thereby enhancing its ability to pick up small biases and more accurately navigate towards the most beneficial target. This property that our model predicts makes the organization of the bifurcations along the trajectories crucially important - beyond determining the overall shape of the trajectories, these also define points in space-time near which we expect the brain can amplify discrimination among options based on even very small differences in their perceived quality. 

When comparing the theoretical trajectories to the experiments \cite{previous}, it was found that the spins have modified interactions that can be mapped to a distortion of the relative angles between targets. Using our MF analysis of the trajectories, we show how increasing the angular distortion leads to bifurcations at smaller relative angles between the targets. By spatially spreading the bifurcations more uniformly the distortion simplifies greatly the resulting trajectories. In many spatial configurations of three targets, distortion reduces the number of bifurcations from an infinite number to three, and in the case of four targets it prevents endless indecision that appears in the formation of loops. We therefore conclude that the value of the angular distortion that was found to apply across taxa \cite{previous,oscar2023simple} can be motivated by allowing animals to move along less complex trajectories, resolving convoluted paths that appear when there are more numerous targets. Our results suggest that nature seems to have chosen this form of interactions, as they allow animals to perform more efficient foraging, uniformly exploring arrays of identical targets.

These results highlight the richness of this spin model, where movement through space is determined by spin-spin interactions, which are in turn dependent on the position (of the animal or group) with respect to the targets. It forms the first step towards further theoretical analysis of this model under more realistic conditions, such as in the presence of bias between the targets. 
The model is also of interest from a broader theoretical physics perspective due to its coupling of equilibrium spin dynamics and propulsion of active-matter particles, as well as its connection to general research on decision-making in moving agents.

\section{Acknowledgments}
N.S.G. is the incumbent of the Lee and William Abramowitz Professorial Chair of Biophysics. This research is made possible in part by the historic generosity of the Harold Perlman Family. D.G. and I.D.C. acknowledge support from the Office of Naval Research Grant N0001419-1-2556, Germany’s Excellence Strategy-EXC 2117-422037984 and the Max Planck Society, Horizon Europe, PathFinder European Innovation Council Work Programme (Grant agreement no. 101098722) as well as the European Union’s Horizon 2020 research and innovation programme under the Marie Skłodowska-Curie Grant agreement (\#860949).

\appendix*
\clearpage
\section{Appendix}

\renewcommand{\theequation}{A.\arabic{equation}}
\setcounter{equation}{0}

\section{The association of an instability (bifurcation) with the existence of at least two stable solutions} 
\label{proposition}
\textbf{Proposition 1} \textit{There is an unstable solution to the model equations at a point $\vec{r}_b$, if and only if there are at least two stable solutions at this point.}

This proposition gives us an alternative definition for the bifurcation point as a point where there exist at least two stable solutions. It is easy to see why it is true if we look at the expression for the effective free energy of the spins (Eq.~(\ref{free})), as a function of $\pm\mid \vec{V}\mid$ (one dimensional projection).  For large values of $\mid \vec{V} \mid$ the free energy diverges
\be \label{limit}
\lim_{\pm\mid \vec{V}\mid \rightarrow \infty}F(\vec{V},T)=\infty.
\ee
An unstable point at $\vec{r}_b$ corresponds to a local maximum of $F(\vec{V},T)$ at $\vec{V}(\vec{r}_b)$. Together with Eq.~($\ref{limit}$) (and Rolle's lemma) we conclude that there should exist at least two minima for the function $F(\vec{V},T)$, that correspond to stable solutions of the model equations. If we start from two stable solutions (minima of $F(\vec{V},T)$) according to Rolle's lemma there should be at least one unstable point between them. $\Box$

\begin{figure}[htb]
\centering
\includegraphics[width=0.8\linewidth]{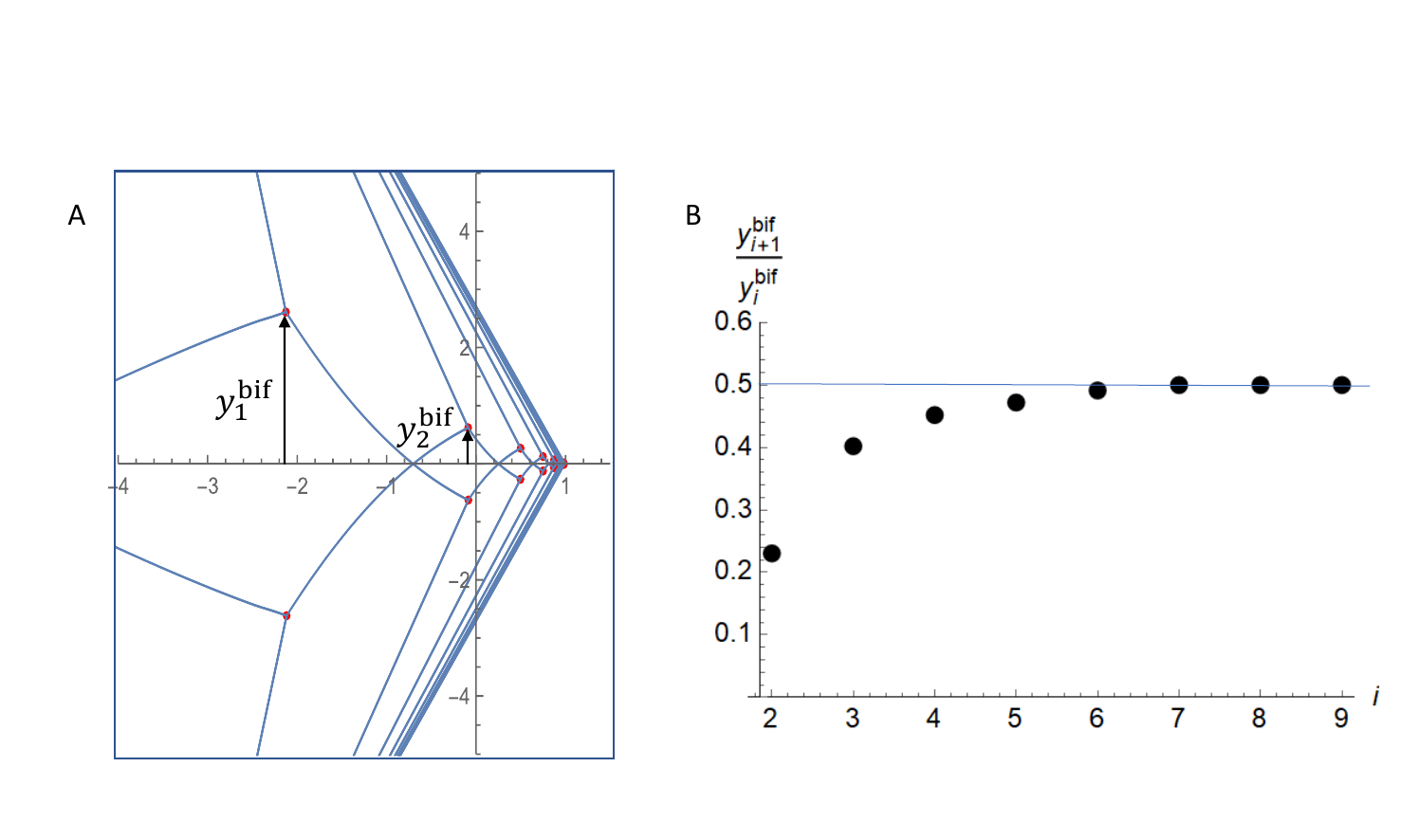}
\caption{\label{Convergence}
A numerical demonstration of the convergence of the self similar pattern towards the central target.
(A) A magnification of the inset from Figure~\ref{BifurCurves}G. We look at the ratios of consecutive distances of the bifurcation points from the symmetry axis $y^{bif}_{i+1}/y^{bif}_{i}$ and show in (B) that the ratio converges to an infinite geometric series with a common ratio of 0.5.
}==
\end{figure}

\section{Alternative way to check stability}
We can write the system of equations (\ref{DynVel}) through its projections on the directions of the targets $\hat{p}_{i}$ ($i=1,..,k$):
\be \label{Pi_eq}
\frac{dV_{p_{i}}}{dt}=\sum_{j=1}^{k}\frac{\bar{v}\hat{p}_j\cdot\hat{p}_i}{k\left(1+\exp\(-\frac{2\,k\,\bar{v}\,V_{p_{j}}}{T}\)\right)}-V_{p_i},
\ee
where $V_{p_i}\equiv \vec{V}\cdot \widehat{p}_i$.

Let $V^{(0)}_{p_i}$ be a solution of equations~(\ref{Pi_eq}) and consider a small perturbation 
\be
V_{p_i}=V^{(0)}_{p_i}+\epsilon_i.
\ee
Then the equations for the linear perturbations (first order in $\epsilon_i$) are
\be \label{Stab_alt}
\frac{d\epsilon_i}{dt}=\sum_{j=1}^{k}T_{ij}\epsilon_j
\ee
where
\be
T_{ij}\equiv\frac{\bar{v}^2\hat{p}_j\cdot\hat{p}_i}{2\,T}\sech^2\(\frac{k\,\bar{v}\,V_{p_j}}{T}\)-\delta_{ij}.
\ee

Then the solution $V^{(0)}_{p_i}$ is stable if and only if $T_{ij}$ is negative-definite. The advantage of this formulation is that it can be extended to the distorted angle $\tilde{\theta}_{ij}$ with $\nu<1$  (Eqs.~(\ref{theta_def}-\ref{dist})).

\subsection{Derivation of the effective free energy of the spins}
From the two-dimensional Hamiltonian ~(Eq.~(\ref{Hamiltonian})) we can obtain the effective free energy in the MF approximation, following the procedure that was used for the one dimensional Curie-Weiss model (see for example~\cite{statistical}, chapter 13), and was applied to the current Hamiltonian in~\cite{PhysRevE.97.032304} for the case of two targets. Here we extend the derivation to $k$ targets.
The starting point is the partition function for the Hamiltonian from Eq.~(\ref{Hamiltonian}), which is
\be
Z=tr_{\{\sigma_i\}}Exp\left[\frac{k\,\bar{v}^2}{N\,T}\(\sum_{i=1}^{N}\sigma_i\,\widehat{p}_i\)^2\right],
\ee
where here $\widehat{p}_i\in\{\widehat{p}_1,...,\widehat{p}_k\}$ and we have in the sum an equal number of spins for every target $N/k$ (assuming $N \gg 1$), while $\bar{v}$ is a constant whose dimension is velocity.
Since the system is two-dimensional, let us introduce two auxiliary fields
\be
\vec{V}\equiv(V_x,V_y)
\ee
and using the Gaussian identity
\be
e^{\frac{a^2+b^2}{4\,c}}=\frac{c}{\pi}\int_{-\infty}^{\infty} dV_{x} dV_{y} \,e^{-c\,(V_{x}^2+V_{y}^2)+a\,V_{x}+b\,V_{y}}
\ee

where we take
\bea
a&=&\frac{2\,k\,\bar{v}}{T}\(\sum_{i=1}^{N}\sigma_i\,\widehat{p}_i\)_x \nonumber\\
b&=&\frac{2\,k\,\bar{v}}{T}\(\sum_{i=1}^{N}\sigma_i\,\widehat{p}_i\)_y \nonumber\\
c&=&\frac{N\,k}{T},
\eea
we can write the partition function in a form which is linear in $\sigma_i$:
\be
Z=\frac{N\,k}{\pi\,T}\,tr_{\{\sigma_i\}}\int_{-\infty}^{\infty}dV_x\,dV_y\,\exp\(-\frac{N\,k}{T}\,\vec{V}^2+\frac{2\,k\,\bar{v}}{T}\,\vec{V}\cdot\sum_{i=1}^{N}\sigma_i\,\widehat{p}_i\).
\ee
Then summing over the possible values of the spins we get
\be
tr_{\{\sigma_i\}}\exp\(\frac{2\,k\,\bar{v}}{T}\,\vec{V}\cdot\sum_{i=1}^{N}\sigma_i\,\widehat{p}_i\)=\prod_{i=1}^{N}\sum_{\sigma=0,1}\exp\(\frac{2\,k\,\bar{v}}{T}\,\vec{V}\cdot\widehat{p}_i\,\sigma\)=\prod_{i=1}^{N}\,e^{\ln\(1+\exp\(\frac{2\,k\,\bar{v}}{T}\,\vec{V}\cdot\widehat{p}_i\)\)}=e^{\frac{N}{k}\sum_{i=1}^{k}\ln\(1+\exp\(\frac{2\,k\,\bar{v}}{T}\,\vec{V}\cdot\widehat{p}_i\)\)},
\ee
when we sum over $N/k$ spins for each target and where the last summation is not over individual spins but over the unit vectors towards the targets $\widehat{p}_i$ ($i=1,..,k$).

Then we can read the free energy per spin $F(\vec{V},T)$ by comparison to the general form
\be
Z \sim \int_{-\infty}^{\infty}dV_x\,dV_y\,\exp\(-\frac{N}{T}\,F(\vec{V},T)\)
\ee


and obtain
\be
F(\vec{V},T)=k\,\vec{V}^2-\frac{T}{k}\sum_{i=1}^{k}\ln\(1+\exp\(\frac{2\,k\,\bar{v}}{T}\,\vec{V}\cdot\widehat{p}_i\)\).
\ee
\section{Trajectory equations for three targets}

For a system at a point $(x_0,y_0)$ and three targets at $(x_i,y_i)$ $i=1,..,3$, we define the following vectors

\be
\widehat{p}_i=\(\frac{x_i-x_0}{\sqrt{(x_i-x_0)^2}+(y_i-y_0)^2},\frac{y_i-y_0}{\sqrt{(x_i-x_0)^2}+(y_i-y_0)^2}\),
\ee
and the proportion of ``on'' spins in each group

\be
n_i=\frac{1}{3\(1+\exp\(-\frac{6\,\bar{v}\,V_{p_i}}{T}\)\)}, \label{ni3}
\ee
where $V_{p_i}\equiv \vec{V}\cdot \widehat{p}_i$.

Then we obtain the following set of equations for $V_{pi}$ (by taking projections of Eq.~(\ref{velocity}) and the definitions in~Eq.~(\ref{dist}):
\bea \label{vp1}
V_{p1}&=&\bar{v}
\(n_1+n_2\cos\tilde{\theta}_{12}+n_3\cos\tilde{\theta}_{13}\)\\
V_{p2}&=&\bar{v}\(n_1\cos\tilde{\theta}_{12}+n_2+n_3\cos\tilde{\theta}_{23}\) \label{vp2}\\
V_{p3}&=&\bar{v}\(n_1\cos\tilde{\theta}_{13}+n_2\cos\tilde{\theta}_{23}+n_3\).\label{vp3}
\eea
Let $\hat{t}^{0}$ be the (tangent) velocity unit vector, where  $\hat{n}^{0}$ is the perpendicular direction. Then since
\be
(\hat{n}^{0}\cdot \hat{p}_i)^2=1-(\hat{t}^{0}\cdot \hat{p}_i)^2
\ee
we get the relation
\be
(\hat{n}^{0}\cdot \hat{p}_i)^2=1-\frac{V_{pi}^2}{\vec{V}^2.}
\ee 
Solving Eqs.~(\ref{vp1}-\ref{vp3}) for $V_{pi}$ and substituting
into Eq.~(\ref{Adefined}) we get the following criterion for stability:
\be \label{stability}
A=1-\frac
{\bar{v}^2}{2\,T}\sum^{3}_{i=1}\sech^2\(\frac{3\,\bar{v}\,V_{pi}}{T}\)\(1-\frac{V_{pi}^2}{\vec{V}^2}\)>0.
\ee

When this condition is violated along the trajectory, we identify a bifurcation point.

\section{MF trajectories at high and low temperatures}

\begin{figure}[htb]
\centering
\includegraphics[width=\linewidth]{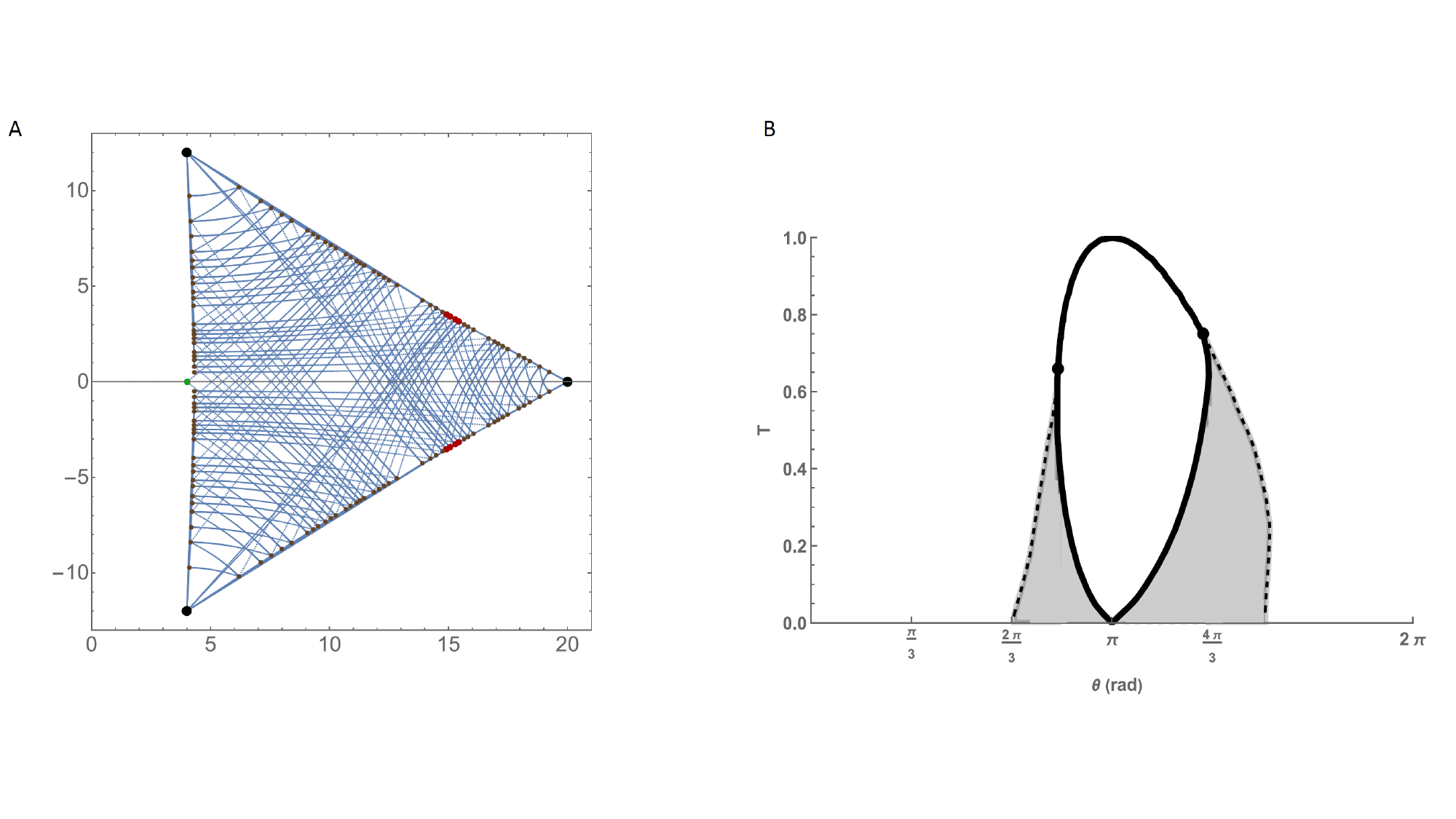}
\caption{\label{Tt0}
(A) The bifurcation pattern at $T\rightarrow 0$. Here $T=10^{-4}$. The targets are at $(4,12)$,$(4,-12)$ and $(20,0)$. Bifurcation depth is 8. The green point is the first bifurcation point and the only one with two outgoing trajectories (binary). The brown points are bifurcation points with three outgoing trajectories and the red points are with five. The bifurcation points accumulate along bifurcation curves that in the limit $T\rightarrow 0$ are straight lines that connect the targets. 
(B) The full phase diagram for three targets includes another phase transition at large angles that could correspond to a movement backwards in an opposite direction to the original one. It shows that the critical angle between the left and right targets is $\Pi$ for $T=0$, which corresponds to the bifurcation curve that connects those targets in A which is a straight line.  
}
\end{figure}

\begin{figure}[htb]
\centering
\includegraphics[width=0.8\linewidth]{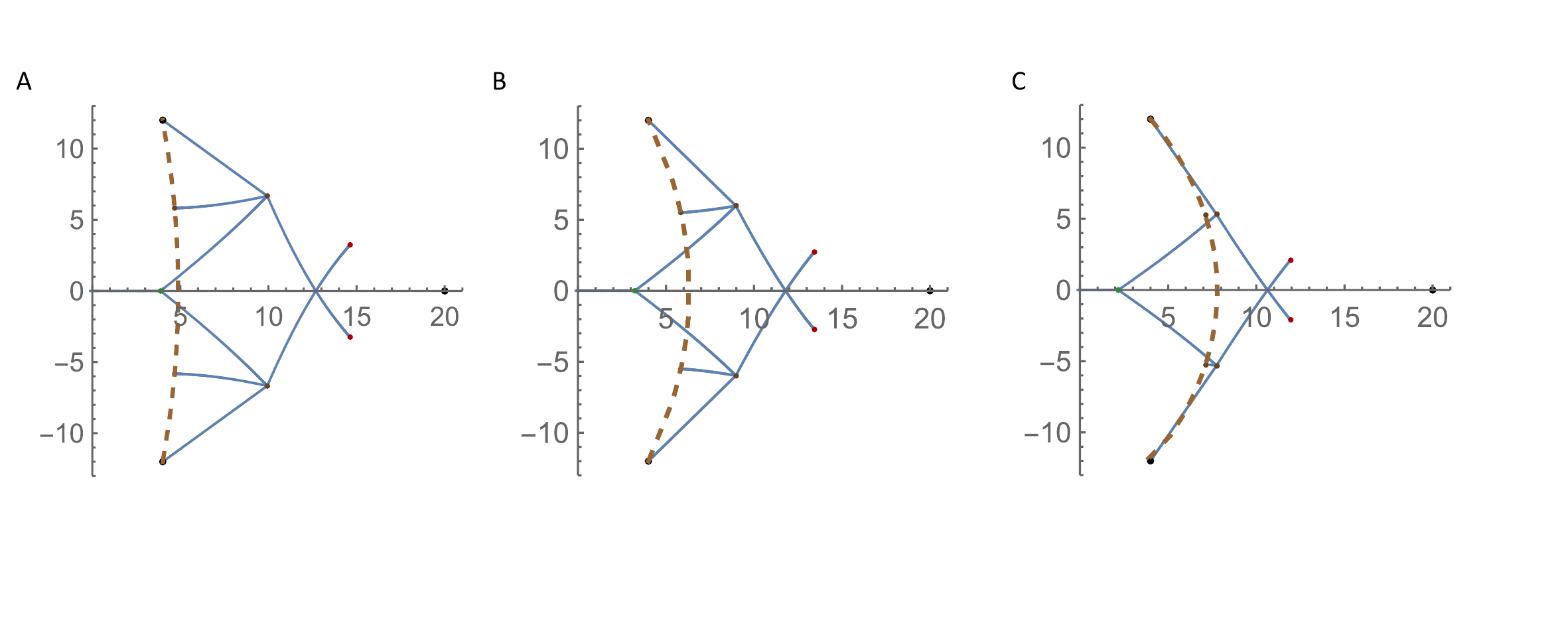}
\caption{\label{SecondBifurcationTgg0}
The second bifurcation at low values of the temperature $T$. The targets and the colors of the bifurcation points are like in Fig.~\ref{Tt0}A. The dashed brown line is the bifurcation curve that connects the left and right targets. In all the cases we see the first bifurcation (with two outgoing trajectories) and the second one (with three outgoing). 
(A)$T=3\cdot10^{-3}$.
(B)$T=3\cdot10^{-2}$.
(C)$T=0.12$.
As the temperature is increased, the bifurcation curve moves to the right according to the phase diagram in~\ref{Tt0}B. As a result, the outgoing trajectory backwards (from the second bifurcation point) becomes shorter
and at high enough temperature, the second bifurcation remains with only two outgoing trajectories. This is the case in all the previous examples with $T=0.2$. 
}
\end{figure}

\begin{figure}[hbt]
\centering
\includegraphics[width=\linewidth,height=0.5\linewidth]{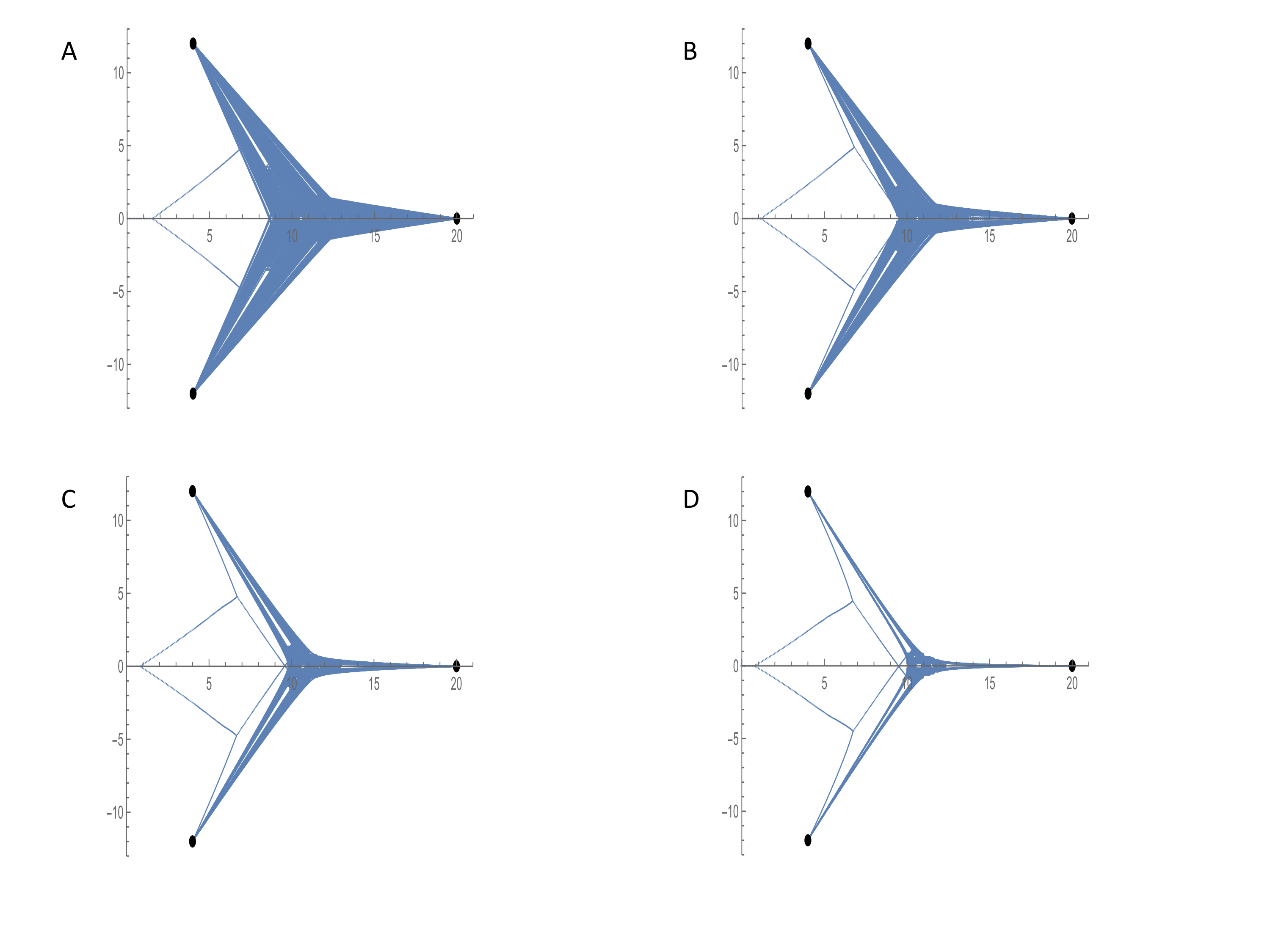}
\caption{\label{HighTexamples}
The pattern of the trajectories for higher values of the temperature $T$. The targets are at $(4,12)$,$(4,-12)$ and $(20,0)$. Bifurcation depth is 12. As the temperature is increased, the bifurcations backwards move more to the right according to the phase diagram~\ref{Tt0}B, and the area of the ``space filling trajectory'' shrinks.
(A) $T=0.2$.
(B) $T=0.3$.
(C) $T=0.4$.
(D) $T=0.5$.
}
\end{figure}

We find that the simplest form of the BC appear in the limit $T\rightarrow0$ (Fig.~\ref{Tt0}A). In this limit the bifurcation curves are (almost) straight lines that connect two of the three targets. The reason for the BC approaching straight lines becomes clear when looking at the phase diagram for movement along the axis of symmetry for three targets (Fig.~\ref{Tt0}B). When the system moves along the axis of symmetry, the bifurcation occurs at the critical angle $\theta_{crit}\rightarrow\pi$, which corresponds to points along the straight lines that connect the corresponding targets. When the temperature is increased the BC acquires curvature and eventually the backwards outgoing trajectories disappear (Fig.~\ref{SecondBifurcationTgg0}). At higher temperature, the ``space filling'' pattern shrinks even more (Fig.~\ref{HighTexamples}).

\section{Loops in trajectories for four targets}
\label{loops}
\begin{figure}
\centering =====--======
\includegraphics[width=\linewidth]{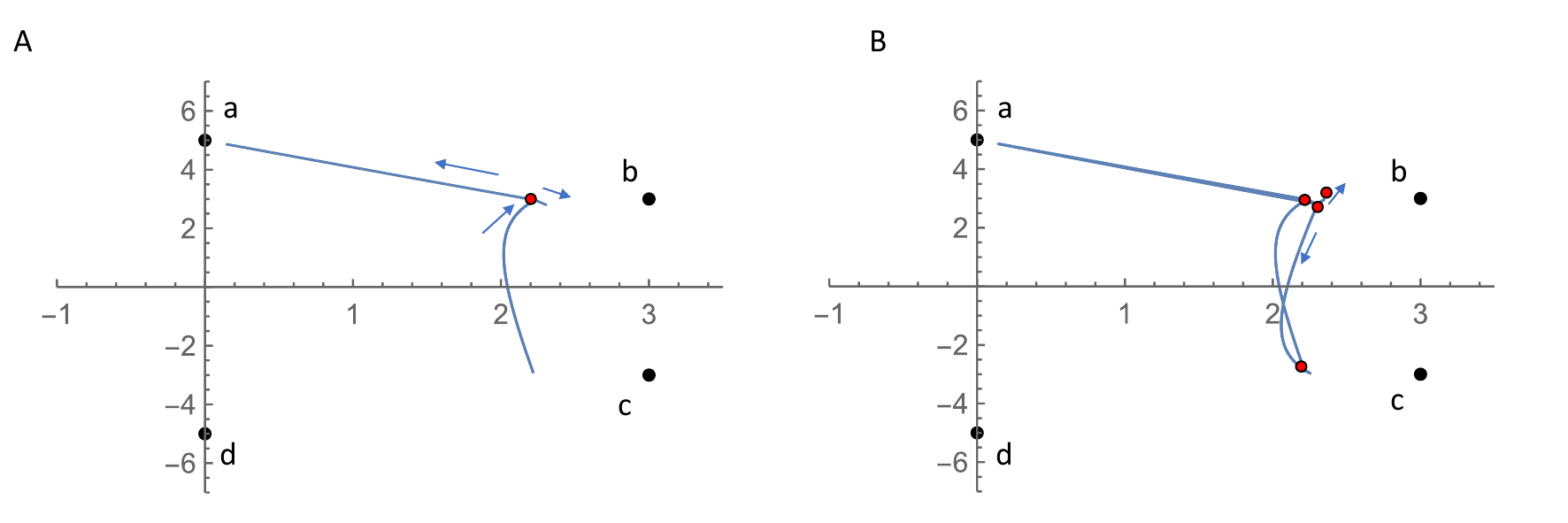}
\caption{The formation of a loop with four targets that are denoted by (a,b,c,d) according to Fig.~\ref{Reducing}E.
(A) Here we have two outgoing trajectories from a bifurcation point (the red dot) Decision towards a or a compromise between c and d (downwards).  (B) Here we see the following three bifurcation points that show a closure of a loop (up to a small shift).} 
\label{loop}
\end{figure}

In order to understand how loops form in trajectories with four targets, let us follow the bifurcation pattern starting from a particular bifurcation point in Fig.~\ref{Reducing}E. In Fig.~\ref{loop}A the bifurcation point is denoted by red dot and the targets by $a,b,c,d$ as seen in Fig.~\ref{loop}. The formation of loops in the case of four targets is similar to the formation of the self similar structure for three targets (Fig.~\ref{BifurCurves}G). The loop is created as a sequence of bifurcation points that alternate between a compromise of the two upper targets (a and b) and a compromise of the two lower targets (c and d). The bifurcation point in the example of Fig.~\ref{loop}A corresponds to a compromise of a and b. There are two outgoing trajectories from this point. One of them corresponds to a decision solution, leading towards target a and terminates there, and one that ends with a new bifurcation point which goes downward and corresponds to a compromise between c and d (the short segment). We can see the new bifurcation point in Fig.~\ref{loop}B. Then, from this point, there are two outgoing trajectories, one which corresponds again to a compromise of a and b (the short segment upwards) and a second one that corresponds to a compromise of c and d. This way a loop is formed (up to a small shift) where in the next step (not shown) we would have a decision towards d or a compromise of a and b.
\clearpage 
\bibliography{SpinModel1.bib}

\end{document}